\documentclass[11pt]{article}
\usepackage{authblk}
\usepackage{fullpage}
\usepackage{amssymb,amsmath,mathtools}
\usepackage[T1]{fontenc}
\usepackage{siunitx}
\usepackage[version=3]{mhchem}
\usepackage{xr}
\usepackage{bm,bbm}
\usepackage{pdfpages}
\usepackage{float}
\usepackage{graphicx}
\usepackage{dcolumn}
\usepackage{bm}
\usepackage{color}
\usepackage{tikz}
\usepackage{setspace}
\usepackage{cancel}
\usepackage[title]{appendix}
\usepackage[center,font=small,labelfont=bf]{caption}
\usepackage{float}
\usepackage{braket}
\usepackage{xcolor}

\usepackage[unicode=true]{hyperref}
\hypersetup{colorlinks=true,
            citecolor=blue,
            linkcolor=blue}
            
\usepackage[square,numbers,sort&compress]{natbib}
\usepackage{setspace}

\numberwithin{equation}{section}

\usepackage[left]{lineno}

\newcommand*\linenomathpatch[1]{%
  \cspreto{#1}{\linenomath}%
  \csappto{end#1}{\endlinenomath}%
  \csappto{end#1*}{\endlinenomath}%
}

\linenomathpatch{equation}
\linenomathpatch{gather}
\linenomathpatch{multline}
\linenomathpatch{align}
\linenomathpatch{alignat}
\linenomathpatch{flalign}

\floatname{algorithm}{Algoritmo}

\makeatletter

\newcommand{\hypgeo}[2]{%
  {\vphantom{F}}_{#1}\kern-\scriptspace F_{#2}%
}

\newcommand\approxsim{\mathchoice
  {\@approxsim {\displaystyle}      {1ex} }
  {\@approxsim {\textstyle}         {1ex} }
  {\@approxsim {\scriptstyle}       {.7ex}}
  {\@approxsim {\scriptscriptstyle} {.5ex}}}
\newcommand\@approxsim[2]{%
  \mathrel{%
    \ooalign{%
      $\m@th#1\sim$\cr
      \hidewidth$\m@th#1.$\hidewidth\cr
      \hidewidth\raise #2 \hbox{$\m@th#1.$}\hidewidth\cr
    }%
  }%
}

\newcommand{\bo}{\raise-1mm\hbox{\Large$\Box$}}

\definecolor{deeppink}{rgb}{1.0, 0.08, 0.58}

\begin{document}

\title{Spatial moment dynamics and biomass density equations provide complementary, yet limited, descriptions of pattern formation in individual-based simulations}

\author[1,$\dag$]{Anudeep Surendran}
\author[1,$\dag$]{David Pinto-Ramos}
\author[1,2,3,$\dag$]{Rafael Menezes}
\author[1,3,*]{Ricardo Martinez-Garcia}
\date{}

\affil[1]{\small Center for Advanced Systems Understanding (CASUS); Helmholtz-Zentrum Dresden-Rossendorf (HZDR), Görlitz, Germany}
\affil[2]{Ecology Department, University of São Paulo, São Paulo SP, Brazil}
\affil[3]{ICTP South American Institute for Fundamental Research \& Instituto de F\'isica Te\'orica, Universidade
Estadual Paulista - UNESP, São Paulo SP, Brazil}
 \affil[$\dag$]{These authors contributed equally to this work and share first authorship}
 \affil[*]{Corresponding author: r.martinez-garcia@hzdr.de}

    \maketitle

    \begin{abstract}
     Spatial patterning is common in ecological systems and has been extensively studied via different modeling approaches. Individual-based models (IBMs) accurately describe nonlinear interactions at the organism level and the stochastic spatial dynamics that drives pattern formation, but their computational cost scales quickly with system complexity, limiting their practical use. Population-level approximations such as spatial moment dynamics (SMD)---which describe the moments of organism distributions---and coarse-grained biomass density models have been developed to address this limitation. However, the extent to which these approximated descriptions accurately capture the spatial patterns and population sizes emerging from individual-level simulations remains an open question. We investigate this issue considering a prototypical population dynamics IBM with long-range dispersal and intraspecific competition, for which we derive both its SMD and coarse-grained density approximations. We systematically compare the performance of these two approximations at predicting IBM population abundances and spatial patterns. Our results highlight that SMD and density-based approximations complement each other by correctly capturing these two population features within different parameter regimes. Importantly, we identify regions of the parameter space in which neither approximation performed well, which should encourage the development of more refined IBM approximation approaches.
    \end{abstract}

\section{Introduction}\label{sec:intro}
Populations of sessile organisms exhibit diverse spatial patterns, from randomly scattered individuals to aggregates that are periodically distributed in space \cite{Pringle2017, Lee2021}. The latter are particularly interesting from a modeling point of view because they raise questions about how interactions operating at the scale of organisms lead to organized structures at much larger spatial scales \cite{Rietkerk2008, Martinez-Garcia2022,Tarnita2024}. Moreover, regular patterns could enhance population productivity, resilience, and resistance to higher environmental stresses, underscoring their ecological significance \cite{Rohani1997, Dong2019, rietkerk2021evasion, Kefi2024}. 

Plants in water-limited ecosystems provide a good system to investigate regular pattern formation in populations of sessile and semi-sessile organisms \cite{Rietkerk2002}. Several studies using aerial and satellite images from different locations have identified a variety of patterns in these systems, including gaps, stripes, rings, labyrinths, and spots \cite{Lefever1997, Deblauwe2011, Clerc2021, Guirado2023}. These patterns are primarily shaped by environmental factors such as gradients in resource availability, interactions between plants and other organisms, and local topography \cite{Gandhi2018, Gowda2018, Siteur2023, Pinto-Ramos2023, Guirado2023, Hidalgo2024}. However, the long timescales governing vegetation dynamics and the limited availability of long-term datasets pose challenges for empirically investigating the ecological significance of vegetation patterns \cite[but see][]{Bastiaansen2018, Veldhuis2022}. Alternatively, researchers have used spatially explicit theoretical models to explain both how patterns could form and their ecological significance 
\cite{Lejeune1999, Meron2004, Kealy2012, MartinezGarcia2013, PintoRamos2024}.

Individual-based models (IBMs), also called interacting particle systems in the physics literature \cite{Durrett1994}, provide a straightforward way to describe vegetation spatial dynamics and study pattern formation \cite{Bolker1997,Wiegand2021}. The simplest IBM formulations describe a population as a spatial distribution of point-like organisms that reproduce and die according to probabilistic rules that encode different interactions and dispersal mechanisms \cite{Bolker1999, Wiegand2021}. These models are easy to simulate computationally and can resolve the structure of spatial patterns at the level of single organisms. However, the computational overhead associated with these simulations scales rapidly with the system size and the complexity of interactions, which has motivated the development of a series of analytical tools to approximate IBMs by deterministic dynamical systems. One of these approximations consists in coarse-graining discrete IBMs, characterizing the population dynamics by a biomass density field that changes in space and time following a partial differential equation (PDE) \cite{MartinezGarcia2023, Borgogno2009}. This approach resolves spatial structures explicitly and can predict the dynamics of certain spatial processes, such as the direction of ``desertification'' fronts \cite{Zelnik2017}. However, density-based models do not resolve individual plants, which limits how much they inform about patch composition, and are accurate only in high-density systems. This limitation makes them inappropriate for studying random and irregular aggregated patterns \cite{law2003} unless they consider environmental heterogeneity \cite{sheffer2012,yizhaq2014,echeverria2023effect,pinto2022vegetation} or demographic fluctuations explicitly \cite{Butler2009,Martinez-Garcia2013a,Dasilva2014,Karig2018}. Alternatively, individual-based stochastic dynamics can be described through a hierarchy of differential equations for the moments of the spatial distribution of organisms. This approach is known as spatial moment dynamics (SMD), and it accurately predicts population densities when dispersal and competition kernels are Gaussian \cite{law2003,Bolker1997,Surendran2020}, even for low-density systems in which the biomass field may not be well defined. However, SMD does not resolve spatial structure directly, as it predicts only spatial averages and any spatial structure must instead be inferred from correlation functions \cite{Bolker1997}.
 
In this study, we systematically compare the performance of spatial moment dynamics and density field approximations at recovering both the population abundances and spatial patterns observed in IBM simulations. We run this comparison across a variety of spatial patterns commonly formed by vegetation in water-limited ecosystems---including homogeneous, segregated, and regular aggregated patterns \cite{MartinezGarcia2023}---by systematically varying dispersal and competition scales and kernels. Our results suggest that density field PDEs are better at reproducing the population size when the population exhibits regular aggregated patterns. In contrast, for segregated spatial patterns---where individuals tend to be isolated and density fields become ill-defined---spatial moment dynamics offer more reliable estimates for the average density and the overall point pattern structure. These findings highlight the complementary strengths of the two approximations, defining their respective regimes of validity in modeling the spatial population dynamics of sessile and semi-sessile organisms.

\section{Model formulation and analytical approximations}\label{sec:description}
\subsection{Description of the individual-level stochastic dynamics}\label{sec:IBM}

We assume that organisms are identical except for their location and consider that they inhabit a two-dimensional landscape $\Omega$, with area $\lvert\Omega\rvert  = A$. Therefore, the state of the population at a given time $t$ is fully determined by the set of plant locations ${\mathbf{x}}_i$ $[i=1,\ldots N(t)]$, where $N(t)$ is the number of plants at time $t$. Altogether, these locations define a spatial pattern $p(\mathbf{x},t)=\sum_i \delta(\mathbf{x}-\mathbf{x}_i)$.

To keep our analysis as simple as possible, we consider a birth-death dynamics in continuous time \cite{Bolker1999,law2003}. The $i$-th individual, located at $\mathbf{x}_i$, reproduces and disperses a new organism at location ${\mathbf{x}}$ at a rate
\begin{equation}\label{B_i(X)}
    B_i(\mathbf{x}) = b \, \phi_{\mathrm{D}}({\mathbf{x}}-{\mathbf{x}}_i),
\end{equation}
where $\phi_{\mathrm{D}}({\mathbf{x}}-{\mathbf{x}}_i)$ is a dispersal kernel that defines the probability density that an offspring establishes at ${\mathbf{x}}$, given the parent is at location ${\mathbf{x}_i}$. Because it acts as a probability density function, $\phi_\mathrm{D}$ must be normalized to unity. Several empirical studies have measured these dispersal kernels across species and investigated their importance in shaping ecological dynamics \cite{Rogers2019,Bullock2017}. Here, we follow a more common choice in the modeling literature and assume this kernel to be a Normal distribution centered at ${\mathbf{x}}_i$ and with variance $s$ \cite{Bolker1999, Surendran2020}. Conversely, each organism in a given spatial distribution $p$ dies at a rate
\begin{equation}\label{D_i(X)}
    D_i(p) = d + g \sum_{j\neq i} \phi_{\mathrm{C}}({\mathbf{x}}_i-{\mathbf{x}}_j),
\end{equation}
where $d$ is the baseline mortality rate, and the second term accounts for density-dependent effects due to intraspecific competition. $\phi_{\mathrm{C}}$ is the competition kernel that weighs the influence of an organism at ${\mathbf{x}}_j$ on the focal individual at ${\mathbf{x}}_i$ \cite{A_Surendran_2020}. The competition kernel models the increased mortality an individual might experience as it competes with neighbors for shared limited resources. The shape of the competition kernel is a key ingredient in determining the type of spatial patterns that can be formed \cite{Lopez2004,Silvano2024}. In the simulations, we used two different types of competition kernels. The first type is a Gaussian kernel with zero mean and standard deviation $r_c$. The second is a top-hat competition kernel
\begin{equation}\label{eq:kernel}
    \phi_{\mathrm{C}}({\mathbf{x}}_i-{\mathbf{x}}_j) =
    \begin{dcases}
        \frac{1}{\pi r_c^2} & \mbox{if} \ \rvert{\mathbf{x}}_i-{\mathbf{x}}_j\rvert\leq r_c, \\
        0 & \mbox{otherwise}. \\
    \end{dcases}
\end{equation}
In both cases, $r_c$ defines the characteristic spatial scale of the competition. 

Note that, while we present a conceptual model, every model parameter quantifies a process driving plant population dynamics in natural scenarios and can thus be linked to different plant traits. The standard deviation of the dispersal kernel, $s$, is a proxy for the distance over which a plant disperses its seeds. The characteristic spatial range of competition, $r_c$, defines the distance over which plants' roots compete for belowground shared resources such as water and minerals. Finally, the intrinsic birth and death rates, $b$ and $d$, represent the rates at which an individual plant reproduces and dies, respectively, while in isolation. Therefore, even though our model does not represent a specific plant species, approximate parameter values can be estimated from empirical measurements of dispersal kernels, root systems, or plant life-history studies \cite{Hirsch_2012, Cabal_2020}.

The combined effect of birth and death terms in Eq.\,\eqref{B_i(X)} and Eq.\,\eqref{D_i(X)} produces a simple logistic-like birth-death dynamics. We performed all IBM simulations in a square domain of lateral length $L=20$, using periodic boundary conditions. We fixed this domain size to ensure that it is sufficiently large compared to the interaction scales $s$ and $r_c$ to observe meaningful patterns within the domain. A summary of key parameters of the model is provided in Table \ref{Table:parameters}.

\begin{table}[H]
	\centering
	\begin{tabular}{|c | c | c |} 
		\hline
		Symbol& Parameter & Value \\ 
		\hline	
		$b$ & Intrinsic birth rate & 3.05  \\ 
		$d$ & Intrinsic death rate & 0.305 \\
		$s$ & Dispersal range & 0 -- 1.25 \\
		$r_c$ & Spatial range of competition & 0 -- 4 \\%
		$g$ & Strength of competition & 1  \\
		\hline
	\end{tabular} 
	\caption{Model parameters and typical values.}        
	\label{Table:parameters}   
\end{table}

\subsection{Spatial moment dynamics equations}\label{subsec:SMD}

The discrete dynamics in Section \ref{sec:IBM} can be approximated at the macroscale in terms of the moments of the population spatial pattern. For the kind of spatial point process considered in the IBM, the first spatial moment, $Z_{1}(t)$, represents the average plant density within the population \cite{Plank2015}. More formally, the first spatial moment for a given vegetation spatial pattern, $p(\mathbf{x},t)$, over a two-dimensional landscape $\Omega$, with area $\lvert\Omega\rvert  = A$ is defined as, 
\begin{equation}
    Z_1(t) = \frac{1}{A} \int_\Omega p(\mathbf{x},t) \mathrm{d}\mathbf{x}. \label{eq:1moment-def}
\end{equation}
The second spatial moment, $Z_{2}(\bm{\xi}, t)$, is the average density of pairs of plants separated by a displacement $\bm{\xi}$ \cite{Surendran2021}, which is formally defined as,
\begin{equation}
    Z_2(\bm{\xi}, t) = \frac{1}{A} \int_\Omega p(\mathbf{x},t) \left(p(\mathbf{x} + \bm{\xi},t) - \delta(\bm{\xi})\right) \mathrm{d}\mathbf{x}. \label{eq:2moment-def}
\end{equation}
Similarly, the third spatial moment, $Z_{3}(\bm{\xi}, \bm{\xi}', t)$, corresponds to the average density of triplets of plants separated by displacements $\bm{\xi}$ and $\bm{\xi}'$, respectively \cite{Surendran2018}. Again, for a given vegetation spatial pattern $p(\mathbf{x},t)$, the third spatial moment is defined as,
\begin{align}
Z_3(\bm{\xi}, \bm{\xi}', t) &= \frac{1}{A} \int_\Omega p(\mathbf{x},t) \left(p(\mathbf{x} + \bm{\xi},t) - \delta(\bm{\xi})\right) \times \label{eq:3moment-def}\\
& \hspace{1.5cm} \left(p(\mathbf{x} + \bm{\xi}',t) - \delta(\bm{\xi}') - \delta(\bm{\xi}' - \bm{\xi})\right) \mathrm{d}\mathbf{x} \, , \nonumber
\end{align}
where the delta functions in Eq.\,\eqref{eq:2moment-def} and Eq.\,\eqref{eq:3moment-def} prevent counting pairs and triplets with repeated plants. Finally, we can obtain the radial correlation function from the definition of the second spatial moment \eqref{eq:2moment-def} as
\begin{equation}
    C(r,t) = \frac{1}{2\pi r Z_1^2(t)} \int Z_2(r,\theta,t)rd\theta.\label{eq:C(r,t)}
\end{equation}

Note that normalizing the correlation function in Eq.\,\eqref{eq:C(r,t)} by $Z_{1}^{2}(t)$--the expected density of pairs of plants for a homogeneous  (alternatively termed uniform or \textit{well-mixed}) pattern--ensures that $C(r, t)=1$ in the absence of spatial patterns (Fig.\,\ref{Fig:IBM-diagrams}A: Circle panel) \cite{A_Surendran_2020}. As such, when $C(r, t)>1$, the measured spatial structure has more pairs separated by a distance $r$ than in a homogeneous population. We refer to these types of spatial patterns as clustered (or \textit{aggregated}) spatial patterns (Fig.\,\ref{Fig:IBM-diagrams}A: Triangle panel showing an \textit{irregular} aggregated pattern); aggregated patterns can as well be \textit{regular} if there exists some order between the clusters' positions. On the other hand, when $C(r, t)<1$, there are fewer plant pairs with a separation distance $r$ than if they were in a spatially random configuration. We refer to these patterns as over-dispersed or segregated spatial patterns (Fig.\,\ref{Fig:IBM-diagrams}A: Diamond panel).

To derive the dynamical equations for the spatial moments, we first need to compute the continuum analog of the discrete neighbor-dependent death rate in Eq.\,\eqref{D_i(X)}. According to the definition in Eq.\,\eqref{D_i(X)}, the contribution of a neighboring individual located at a displacement $\bm{\xi}$, to the death rate of a focal plant is given by the competition kernel, $\phi_{C}(\bm{\xi})$. Now, the probability of an individual having a neighbor at a displacement, $\bm{\xi}$, is given by $Z_{2}(\bm{\xi}, t)/Z_{1}(t)$ (see \cite{Surendran2018, Surendran2019} for detailed derivations). Consequently, the expected per-capita death rate can be obtained by multiplying $Z_{2}(\bm{\xi}, t)/Z_{1}(t)$ by $g\phi_{C}(\bm{\xi})$ and integrating over all possible displacements as,
\begin{equation}\label{eq:D_1(t)}
    D_{1}(t)=d+\frac{g}{Z_{1}(t)} \int \phi_{\mathrm{C}}(\bm{\xi})Z_{2}(\bm{\xi}, t)d\bm{\xi}.
\end{equation}
Since birth events are assumed to be neighbor-independent in the IBM [Eq.\,\eqref{B_i(X)}], the expected per-capita birth rate is simply $B_{1}(t)=b$. 

The dynamics of the first spatial moment is entirely governed by the balance between birth and death events because these are the only processes influencing population size in our model. Therefore, the time evolution of the first spatial moment is:
\begin{equation}\label{eq:DynamicsZ_1(t)}
    \frac{d}{dt}Z_{1}(t)=\Big(b-D_{1}(t)\Big) \,Z_{1}(t).
\end{equation}
Note that the dynamics of the first moment, in turn, is governed by the second moment through the death rate term in Eq.\,\eqref{eq:D_1(t)}. 

Next, we derive the dynamics of the second spatial moment. To this end, we first compute the death rate, $D_{2}(\bm{\xi}, t)$, for a plant that is paired with another at a displacement $\bm{\xi}$. Remember that the neighbor-dependent component of the death rate for a single plant, $D_{1}(t)$, was conditioned on the presence of another plant at a displacement $\bm{\xi}$. In a similar fashion, the neighbor-dependent component of $D_{2}(\bm{\xi}, t)$, is conditioned on the presence of a third plant at a displacement $\bm{\xi}'$. This conditional probability of having another plant located at a displacement $\bm{\xi}'$ is $Z_{3}(\bm{\xi}, \bm{\xi}', t)/Z_{2}(\bm{\xi}, t)$. Following the same procedure as in Eq.\,\eqref{eq:D_1(t)}, the expected death rate of a plant paired with another at a separation displacement $\bm{\xi}$ can be computed by multiplying $Z_{3}(\bm{\xi}, \bm{\xi}', t)/Z_{2}(\bm{\xi}, t)$ by $g\phi_{C}(\bm{\xi})$ and integrating over all possible displacements as follows:
\begin{equation}\label{eq:D_2(xi, t)}
    D_{2}(\bm{\xi}, t)=d+\frac{g}{Z_{2}(\bm{\xi}, t)} \int \phi_{\mathrm{C}}(\bm{\xi}')\,Z_{3}(\bm{\xi}, \bm{\xi}', t)d\bm{\xi}'+g\phi_{\mathrm{C}}(\bm{\xi}).
\end{equation}
Note that the additional factor of $g\phi_{C}(\bm{\xi})$ in the third term of Eq.\,\eqref{eq:D_2(xi, t)} accounts for the direct influence of the individual at displacement $\bm{\xi}$.

We compute the dynamics of the second spatial moment by examining how the density of pairs at displacement $\bm{\xi}$ changes, balancing pair loss through death and creation through birth. An existing pair of individuals separated by a displacement $\bm{\xi}$ can be lost if either one of the individuals dies, which occurs at a rate $D_{2}(\bm{\xi}, t)$. Conversely, a new pair separated by $\bm{\xi}$ can be generated in two ways. First, if a pair already exists at displacement $\bm{\xi} + \bm{\xi}'$, one of these individuals can disperse a daughter organism at a displacement $\bm{\xi}'$, forming a new pair with displacement $\bm{\xi}$. This occurs at a rate $b\phi_D(\bm{\xi}')$. Second, a new pair at displacement $\bm{\xi}$ can be created when a single individual disperses an offspring at displacement $-\bm{\xi}$, occurring at a rate $b\phi_D(-\bm{\xi})$. Combining these possibilities, we write the time evolution of the second spatial moment as,
\begin{equation}\label{eq:DynamicsZ_2(xi, t)}
    \frac{d}{dt}Z_{2}(\bm{\xi}, t)=-2D_{2}(\bm{\xi}, t)Z_{2}(\bm{\xi}, t)+2b\int\phi_{D}(\bm{\xi}')Z_{2}(\bm{\xi}+\bm{\xi}', t)\,d\bm{\xi}'+2b\phi_{D}(-\bm{\xi})Z_{1}(t).
\end{equation}
The dynamics of the second moment still depend on the third moment through the definition of $D_{2}$ in \eqref{eq:D_2(xi, t)}, requiring a moment closure approximation to completely solve Eq.\,\eqref{eq:DynamicsZ_2(xi, t)}. Here, we employ the power-2 asymmetric closure:
\begin{equation}\label{eq:P2AClosure}
    Z_{3}(\bm{\xi}, \bm{\xi}', t)=\frac{\Big(4Z_{2}(\bm{\xi}, t)Z_{2}(\bm{\xi}', t)+Z_{2}(\bm{\xi}, t)Z_{2}(\bm{\xi}'-\bm{\xi}, t)+Z_{2}(\bm{\xi}', t)Z_{2}(\bm{\xi}'-\bm{\xi}, t)-Z_{1}^{4}(t)\Big)}{5Z_{1}(t)},
\end{equation}
which approximates the third moment in terms of the first two spatial moments \cite{Murrell_et_al_2004,Surendran2021}.

Moment closures approximating the third moment in terms of the first two moments inherently rely on two key assumptions. First, the third moment must be positive, as it represents triplet density and is calculated from the product of three non-negative densities [Eq.\,\eqref{eq:3moment-def}]. Therefore, the closure approximation must also remain non-negative. The second assumption is the dynamical invariance under relabeling, implying that relabeling identical individuals should not alter the system dynamics \cite{Murrell_et_al_2004}. Different closure methods balance these assumptions, with the power-2 closure we used in our computations offering an optimal trade-off between them. However, in certain parameter regimes, these assumptions may not hold, leading to notable discrepancies between IBM and SMD \cite{Raghib_et_al_2011, Murrell_et_al_2004}. Additionally, truncating the hierarchy of spatial moments at the second order may limit model accuracy for spatial structures with significant amounts of information in spatial moments at higher orders. Nevertheless, many previous studies successfully implemented moment closures to approximate spatial patterns and population dynamics in interacting particle systems \cite{Binny2016, Law_and_Dieckmann_2000, Surendran2020}.

\subsection{Field equation for the biomass density field}\label{sec:PDE}
Additionally, we can develop a coarse-grained equation for the biomass density field. This approximation is particularly useful when the spatial logistic model exhibits regular aggregated patterns, also called periodic spatial patterns. To derive this equation, we first discretize the two-dimensional square domain of lateral length $L$ used in the IBM definition into $m$ square cells of lateral length $\mathrm{d}x=L/m$. Each cell can be labeled with a pair of indices, $(i,j)$ where $i,j=1,..., m$, represent the cell coordinates in the discrete grid. 

In this discretized space, the system's state is given by the number of individuals in each grid cell $\mathbf{\Omega} \equiv \{ N_{11}, N_{12}, ..., N_{mm}\}$. At a given time $t$, the population is found in one of these states with a probability $P(\mathbf{\Omega},t)$ that can be used to compute the expected number of individuals in a focal cell
\begin{equation}
\langle N_{ij} \rangle = \sum_\mathbf{\Omega} N_{ij} P(\mathbf{\Omega}, t),
\end{equation}
where the average is over an ensemble of independent realizations of the individual-based stochastic dynamics. We can formally define the dynamics of this expected occupation number as
\begin{equation}
    \frac{ d \langle N_{ij}\rangle}{d t}  = \sum_\mathbf{\Omega} N_{ij} \frac{\partial P(\mathbf{\Omega}, t)}{\partial t}, \label{DFE}
\end{equation}
where the time derivative of $P(\mathbf{\Omega},t)$ will be given by a master equation of the form
\begin{equation}\label{eq:master}
    \frac{\partial P(\mathbf{\Omega},t)}{\partial t} = \sum_{\mathbf{\Omega}'} {\cal{T}}(\mathbf{\Omega}'\rightarrow \mathbf{\Omega}) P(\mathbf{\Omega}',t) - {\cal{T}}(\mathbf{\Omega}\rightarrow \mathbf{\Omega}') P(\mathbf{\Omega},t)
\end{equation}
where the sum in $\mathbf{\Omega}'$ runs over all possible states and ${\cal{T}}$ are the transition rates between states and are given by the rules of the individual-based stochastic model. Because the probability of simultaneous birth and death events is negligible, the IBM stochastic dynamics is a \textit{one step process} \cite{van1992stochastic}. In terms of the Master equation \eqref{eq:master}, this means that the transition rates $\cal{T}$ between pairs of states that differ in more than one particle are all zero. With this condition and using step operators \cite{Toral2014}, we can write the Master equation as 
\begin{equation}\label{eq:master-step}
    \frac{\partial P(\mathbf{\Omega},t)}{\partial t} = \sum_{kl} \left(E_{kl}^- - 1\right)  {\cal{T}}(\mathbf{\Omega}\rightarrow \mathbf{\Omega}_{kl}^+) P(\mathbf{\Omega},t) + \left(E_{kl}^+ - 1\right)  {\cal{T}}(\mathbf{\Omega}\rightarrow \mathbf{\Omega}_{kl}^-) P(\mathbf{\Omega},t),
\end{equation}
where we have defined the up-down step operators such that
\begin{equation}
    E_{ij}^{\pm}[f(\mathbf{\Omega})] = f[\left\lbrace N_{11},..., N_{ij} \pm 1,..., N_{mm}\right\rbrace]
\end{equation}
and the states $\mathbf{\Omega}_{kl}^{\pm}$ are identical to $\mathbf{\Omega}$ but have $\pm 1$ individual in the $kl$-th element. To fully describe the stochastic dynamics, we need to define the transition rates,
\begin{eqnarray}
{\cal{T}}(\mathbf{\Omega}\rightarrow \mathbf{\Omega}_{kl}^+) &=& \sum_{ij} b\,N_{ij} \phi_\mathrm{D}(\rvert \mathbf{x}_{ij}- \mathbf{x}_{kl} \rvert ) \mathrm{d}x^2, \label{eq:ratebirth} \\
{\cal{T}}(\mathbf{\Omega}\rightarrow \mathbf{\Omega}_{kl}^-) &=& N_{kl} \left( d +  g \sum_{{ij}} N_{ij}\phi_\mathrm{C}(\rvert\mathbf{x}_{ij}-\mathbf{x}_{kl}\rvert)\right). \label{eq:ratedeath}
\end{eqnarray}

Inserting the Master equation \eqref{eq:master-step} with the transition rates in Eqs.\,\eqref{eq:ratebirth}-\eqref{eq:ratedeath} into the definition for the time derivative of the mean number of particles in each spatial cell, Eq.\,\eqref{DFE}, we obtain
\begin{equation}\label{eq:partcorr}
\frac{ d\langle N_{ij} \rangle}{dt}  = b \sum_{kl} \phi_D(\rvert \mathbf{x}-\mathbf{x}'\rvert) \mathrm{d}x^2 \langle N_{kl} \rangle -d\langle N_{ij} \rangle- g \sum_{kl} \phi_C(\rvert \mathbf{x}-\mathbf{x}'\rvert) \langle N_{ij}N_{kl} \rangle,
\end{equation}
where we have used the constraint that $\langle N_{ij} \rangle \geq 0$ $\forall ij$ and thus the probability of states with a negative number of particles, and the transition rates into them, must be zero. Finally, defining a scaled density of individuals, $n(\mathbf{x}_{ij})=N_{ij}g\,/\,b\mathrm{d}x^2$, and time, $\tau = b t$, we obtain a partial differential equation for a population density field  
\begin{equation}\label{eq:denscorr}
\frac{ \partial \langle n \rangle}{\partial \tau} (\mathbf{x}, \tau)  = \int \phi_D(\rvert \mathbf{x}-\mathbf{x}'\rvert) \langle n(\mathbf{x'}) \rangle \mathrm{d}\mathbf{x'} -\frac{d}{b}\langle n(\mathbf{x})\rangle- \int \phi_C(\rvert \mathbf{x}-\mathbf{x}'\rvert) \langle n(\mathbf{x}) n(\mathbf{x'}) \rangle \mathrm{d}\mathbf{x'},
\end{equation}
where we have omitted the time dependence on the right side. We now define the total number of plants in the population as $N=\frac{b}{g}\int\langle n(\mathbf{x})\rangle\,\mathrm{d}\mathbf{x}=Z_{1}\times L^{2}$. For a homogeneous (or well-mixed) population, the total number of plants in the population is denoted by $N_\mathrm{h}=(b-d)/g$. Notice that, similarly to what we obtained in the spatial moment dynamics equations in Section \ref{subsec:SMD}, Eq.\,\eqref{eq:denscorr} does not provide a closed description for the dynamics of the population density, as it also includes its spatial covariance. We could write an equation for the dynamics of this spatial covariance and perform a moment closure approximation, as we did with the spatial moment dynamics equations. This approach is now much more complicated because the plant density is a field rather than a scalar quantity obtained from a spatial average. Thus, a closure approximation is needed to obtain a closed equation for the average density field.

The simplest and most common approximation to treat spatial correlations in Eq.\,\eqref{eq:denscorr} is to neglect the spatial covariance in the density field,
\begin{equation}\label{eq:defcov}
\text{Cov}(\mathbf{x},\mathbf{x}') = \langle n(\mathbf{x})n(\mathbf{x'})\rangle - \langle n(\mathbf{x})\rangle \langle n(\mathbf{x'})\rangle = 0
\end{equation}
which implies that
\begin{equation}\label{eq:mean-field}
\langle n(\mathbf{x})n(\mathbf{x'})\rangle = \langle n(\mathbf{x})\rangle \langle n(\mathbf{x'})\rangle.
\end{equation}
Replacing Eq.\,\eqref{eq:mean-field} into \eqref{eq:denscorr}, we obtain
\begin{equation}\label{eq:PDE-mf}
\frac{ \partial \langle n \rangle}{\partial \tau} (\mathbf{x}, \tau)   = \int \phi_D(\rvert \mathbf{x}-\mathbf{x}'\rvert) \langle n(\mathbf{x'}) \rangle \mathrm{d}\mathbf{x'} -\frac{d}{b}\langle n(\mathbf{x})\rangle- \langle n(\mathbf{x})\rangle \int \phi_C(\rvert \mathbf{x}-\mathbf{x}'\rvert) \langle n(\mathbf{x'})\rangle \mathrm{d}\mathbf{x'},
\end{equation}
which is an extension of the Fisher-Kolmogorov equation with nonlocalities in the competition and growth terms. Defining a change of variables $\mathbf{y} \equiv \mathbf{x}-\mathbf{x}'$ in the first integral of Eq.\,\eqref{eq:PDE-mf} and expanding the density $\langle n(\mathbf{x}-\mathbf{y}) \rangle$ about $\mathbf{x}$ as a Taylor expansion up to second order, Eq.\,\eqref{eq:PDE-mf} reduces to 
\begin{equation}\label{eq:nlFKP}
\frac{ \partial \langle n \rangle}{\partial \tau} (\mathbf{x}, \tau)   = \langle n(\mathbf{x}) \rangle -\frac{d}{b}\langle n(\mathbf{x})\rangle- \langle n(\mathbf{x})\rangle \int \phi_C(\rvert \mathbf{x}-\mathbf{x}'\rvert) \langle n(\mathbf{x'})\rangle \mathrm{d}\mathbf{x'} + \frac{\omega_2}{2} \frac{\partial^2}{\partial x^2} \langle n(\mathbf{x}) \rangle,
\end{equation}
which is the nonlocal Fisher KPP (Kolmogorov–Petrovsky–Piskunov) equation with a diffusion coefficient defined by the reproduction rate (here absorbed in the scaled time) and the second moment of the dispersal kernel, $\omega_2$ \cite{Murray2007ch11}. However, because we are interested in how long-range dispersal influences spatial patterns, we will retain the nonlocal dispersal term.

\section{Results and Discussion}
\subsection{Spatial dynamics with Gaussian kernels}\label{sec:ibm-results}

We first review the behavior of the spatially explicit logistic model presented in Section \ref{sec:IBM} when both the dispersal and competition kernels are Gaussian \cite{law2003,Bolker1997}.
Holding the demographic parameters, $b$, $d$, and $g$ constant, the spatial correlations in the distribution of organisms are controlled by the dispersal and competition range, $s$ and $r_c$ respectively. 

\begin{figure}[ht!]
	\centering
\includegraphics[width=.6\textwidth]{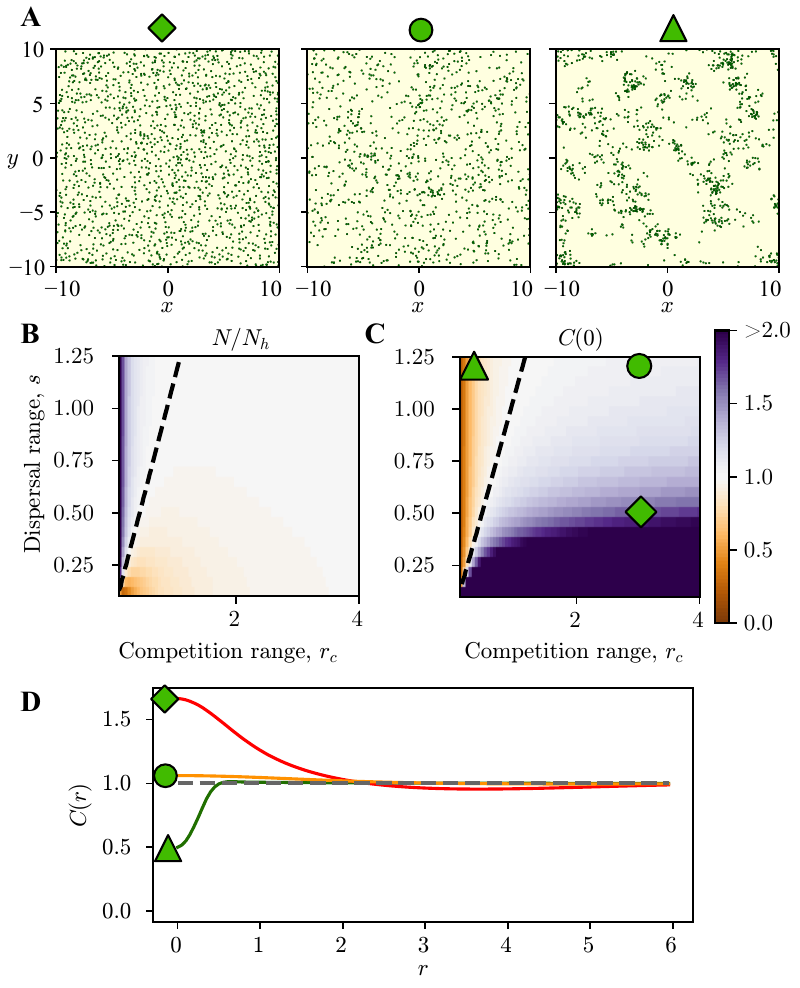}
	\caption{IBM analyses using a Gaussian competition kernel. A) Snapshots of spatial patterns from IBM simulations. From left to right: segregated, homogeneous, and clustered patterns, denoted by diamond, circle, and triangle shapes, respectively. B) Total number of plants normalized by the homogeneous (mean-field) expectation, $N_{\mathrm{h}} = (b-d)/g$. The black dashed line in B correspond to the cases where $N=N_h$ C) Radial correlation function at displacement zero $C(0)$ \cite{law2003} as a function of the dispersal and competition ranges, $s$ and $r_c$ respectively. In both these heatmaps, we plot the steady-state values of $N/N_\mathrm{h}$ and $C(0)$, respectively, and depict time-averaged values computed up to $t=250$. D) Radial correlation function, $C(r)$, as a function of the separation distance, $r$,  for three combinations of $s$ and $r_c$. These parameter combinations are indicated by the coordinates of the green symbols in the heat map and correspond to the respective patterns in A). The black and gray dashed lines  in C and D represent $C(r)=1$.}
	\label{Fig:IBM-diagrams}
\end{figure}

Previous work has identified four possible types of spatial patterns depending on how these two scales compare to each other \cite{law2003}. When both ranges are large, the dynamics are effectively non-spatial with a homogeneous spatial pattern, and the IBM converges to the classical logistic model. However, when the dispersal scale decreases, while keeping competition long-ranged, individuals tend to clump together forming a clustered spatial pattern. This scenario leads to smaller population sizes than in the non-spatial case. Likewise, when the competition range decreases while keeping dispersal long-range, individuals tend to overdisperse, forming a segregated pattern, which increases the total population size. We show the spatial snapshot of these different possible patterns in Fig.\,\ref{Fig:IBM-diagrams}A. Finally, when both competition and dispersal are short-range, newborns cannot escape from competition with their ancestors and the population goes extinct. We recover these results integrating the SMD approximation in Eqs.\,\eqref{eq:DynamicsZ_1(t)}-\eqref{eq:DynamicsZ_2(xi, t)} with the moment closure in Eq.\,\eqref{eq:P2AClosure} (Fig.\,\ref{Fig:IBM-diagrams}B,\,C). In none of the scenarios discussed above, however, the spatial patterns are periodic (see radial correlation functions in Fig.\,\ref{Fig:IBM-diagrams}D).The density field approximation always gives homogeneous solutions in this regime, because, as we mentioned before, it cannot reproduce these irregular structures unless demographic fluctuations or spatial heterogeneities are considered.

\subsection{Spatial dynamics with top-hat kernels: the emergence of periodic patterns}

\begin{figure}[ht!]
	\centering
\includegraphics[width=0.8\textwidth]{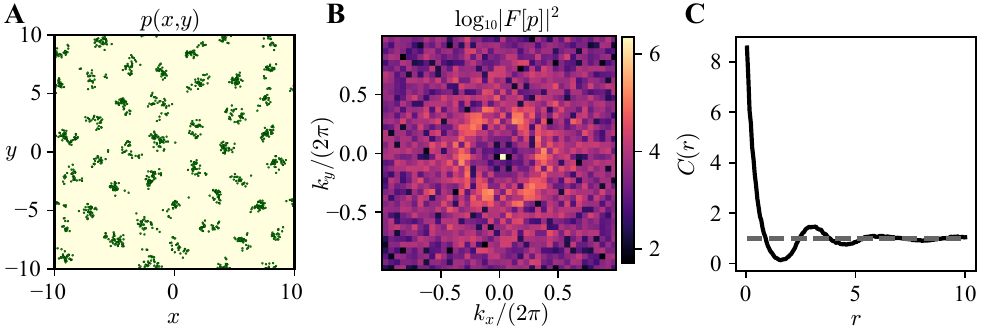}
	\caption{A) Long-time pattern snapshot with dispersal range $s=0.15$ and competition range $r_c=2.2$ (inside the region of high aggregation (labeled with the diamond marker) in the $C(0)$ heatmaps in Fig.\,\ref{Fig:IBM-tophat}). B) Fourier transform of the pattern snapshot. C) Radial correlation function, $C(r)$, of the pattern snapshot.}
	\label{Fig:IBM-pattern}
\end{figure}

The individual-based dynamics defined by the demographic rates in Eqs.\,\eqref{B_i(X)}-\eqref{D_i(X)} can result in periodically arranged clusters of individuals (that is, a regular aggregated pattern) when the competition kernel follows a generalized normal distribution $\phi_\mathrm{C}(\bm{\xi})\propto \exp(-\bm{\xi}/R)^p$ with $p>2$ \cite{Lopez2004}. We performed numerical simulations of the individual-based dynamics using the top-hat kernel from Eq.\,\eqref{eq:kernel}, which corresponds to a generalized normal distribution with $p\rightarrow\infty$. For this kernel choice, we find periodic patterns for short-range dispersal and long-range competition (Fig.\,\ref{Fig:IBM-pattern}A). These periodic patterns can be characterized by their Fourier transform, which will display a circular region with higher power that corresponds to the inverse of the distance between clusters (Fig.\,\ref{Fig:IBM-pattern}B), and by their pair correlation function, which will now oscillate with a frequency given by the inter-cluster distance (Fig.\,\ref{Fig:IBM-pattern}C).

\subsubsection{Accuracy of the spatial moment dynamics approximation}

\begin{figure}[ht!]
	\centering
\includegraphics[width=0.8\textwidth]{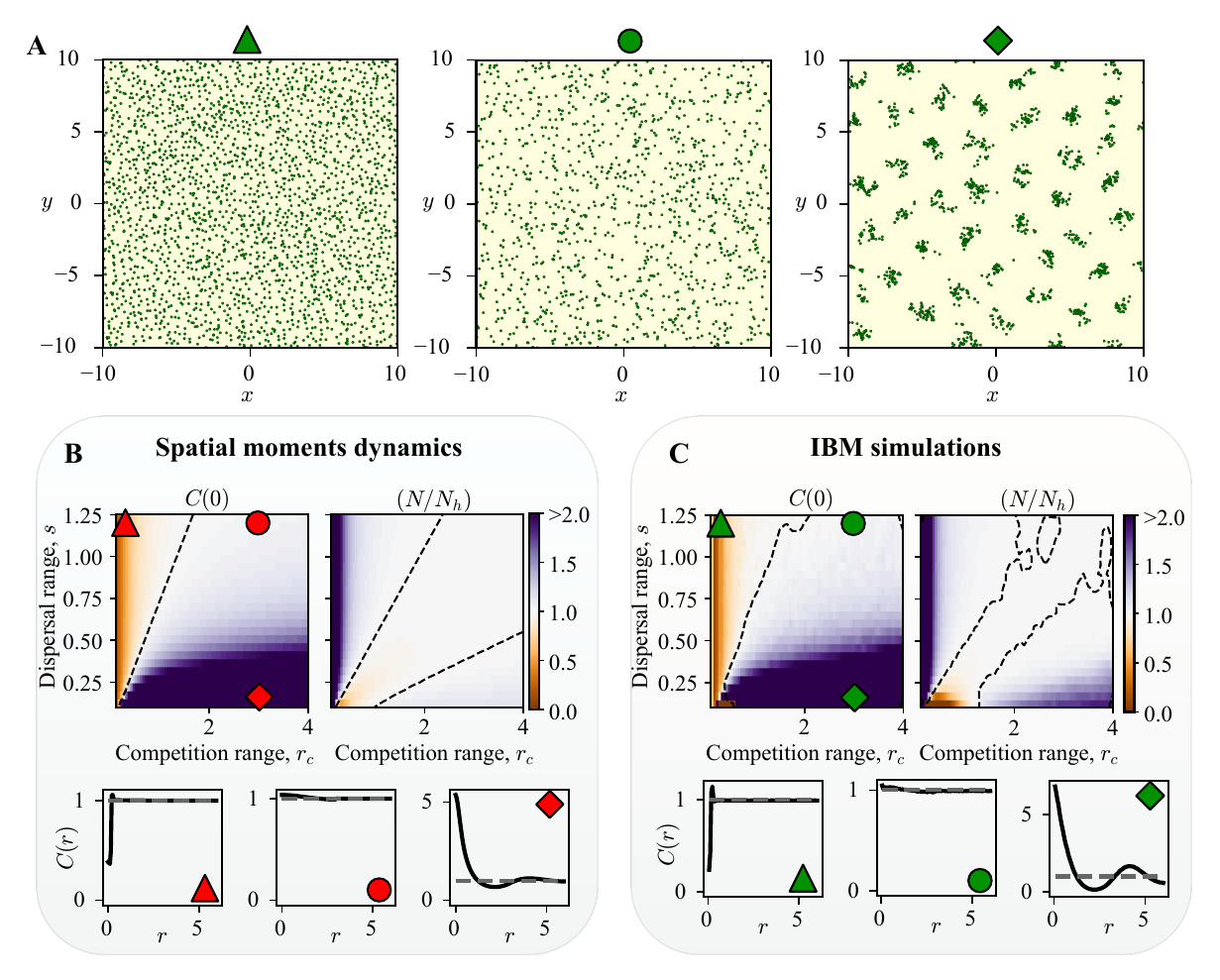}
\caption{Accuracy of SMD approximation reproducing IBM dynamics with a top-hat competition kernel. A) Snapshots of the spatial patterns from IBM simulations. From left to right: segregated, homogeneous, and regular aggregated patterns, denoted by triangle, circle, and diamond shapes, respectively. B) Top panels: SMD predictions for the radial correlation function at zero displacement, $C(0)$, and the total number of individuals normalized by the homogeneous expectation, $N/N_\mathrm{h}$. In both these heatmaps, we plot the steady-state values of $C(0)$ and $N/N_\mathrm{h}$, respectively, which we computed using time-averaged values up to $t=250$. The black dashed line in $C(0)$ plot correspond to $C(0)=1$ and the black dashed lines in $N/N_{h}$ correspond to the cases where $N=N_h$, respectively. The bottom three panels in B show the radial correlation function, $C(r)$, as a function of the separation distance $r$ for three specific combinations of $s$ and $r_c$. These parameter combinations are indicated by the coordinates of the red symbols in the heat maps and correspond to the respective patterns in A). The gray dashed lines represent $C(r)=1$. C) Same as in B but computed directly from stochastic IBM simulations instead of SMD predictions.}
\label{Fig:IBM-tophat}
\end{figure}

To test how accurately SMD equations describe the formation of periodic patterns in the IBM, we solved the SMD Eqs.\,\eqref{eq:DynamicsZ_1(t)}-\eqref{eq:DynamicsZ_2(xi, t)} numerically using a top-hat kernel. SMD equations accurately predict the value of the radial correlation function at pairwise distance zero (top panels in Fig.\,\ref{Fig:IBM-tophat}B,\,C), which is a proxy for local crowding. The SMD equations, however, predict population abundance less accurately. Specifically, when the competition range is large and the dispersal range is short, the SMD equations underestimate the total population size resulting from the IBM simulations (Fig.\,\ref{Fig:IBM-tophat}B,\,C). 

This discrepancy in the total population size comes from the inaccurate prediction of the full radial correlation function provided by the SMD equations. In particular, SMD does not fully capture the amplitude of the oscillations in the radial correlation functions measured from IBM simulations (Diamond panel in the bottom row of Fig.\,\ref{Fig:IBM-tophat}B,\,C). This amplitude, especially when the radial correlation function becomes less than unity, is critical to determine how larger the population size grows when regular patterns form. Smaller values of $C(r)$ indicate that the regions between clusters are less populated, which favors the growth of more individuals within each aggregate \cite{Jorge2024}. 

\subsubsection{Accuracy of the biomass density equation}

The biomass density field equation does not describe the formation of non-periodic spatial patterns, but it can exhibit a symmetry-breaking instability that destabilizes the homogeneous solution and leads to periodic density solutions \cite{Lefever1997,Borgogno2009}. Therefore, we tested how accurately Eq.\,\eqref{eq:PDE-mf} describes the individual-based simulations in the parameter regime where these show periodic patterns. We obtained the onset of pattern formation by performing a linear stability analysis around the non-trivial homogeneous solution
\begin{equation}
\langle n \rangle_h = 1-\frac{d}{b},
\end{equation}
which in the SMD approximation corresponds to imposing $Z_2=Z_1^2$. The linear stability analysis is a standard technique in pattern formation theory that consists on adding small spatiotemporal perturbations to a homogeneous solution of the density equation and calculating the perturbation growth rate. If this growth rate is positive, the perturbation will grow in space and time and could lead to spatial patterns. If perturbations tend to decay, then the uniform solution is stable and patterns do not form \cite{meron2015nonlinear}.
We consider perturbations of the form $\langle n(\mathbf{x})\rangle = \langle n\rangle_h + \Delta n \, e^{i \mathbf{k}\cdot \mathbf{x} + \lambda(\mathbf{k})t}$, where $\Delta n$ is the perturbation amplitude. Inserting this perturbation ansatz in Eq.\,\eqref{eq:PDE-mf} and linearizing around the homogeneous equilibrium, we obtain a perturbation growth rate,
\begin{equation}\label{lin-growth rate}
\lambda(\mathbf{k}) = F[{\phi}_D](\mathbf{k}) -\langle n \rangle_h\,F\left[\phi_C\right](\mathbf{k})- 1,
\end{equation}
where $F[]$ indicates a Fourier transform operation. The onset of pattern formation is defined by the parameter values such that $\lambda(\mathbf{k}_c)=0$, where $\mathbf{k}_c$ is the non-zero Fourier mode at which the perturbation growth rate is maximum, and we can define the parameter combinations where patterns might form by scanning the ($s, r_c$) parameter space and testing for $\lambda(\mathbf{k}_c)\geq 0$.

To quantify how accurately this linear stability analysis predicts periodic pattern formation in the IBM stochastic simulations, we first need to define a metric to characterize the regularity of the stochastic patterns obtained from the IBM simulations. We introduce two different Fourier-based metrics. Because the system is isotropic, we use the radially-averaged Fourier power spectrum, which we obtain after marginalizing the two-dimensional Fourier power spectrum over the polar angle,
\begin{equation}
    S(k)= \frac{1}{2\pi k} \int_0^{2\pi} \lvert F[n](\mathbf{k})\rvert ^2 kd\theta,
\end{equation}
where $n$ is the formal density field $n(\mathbf{r})=\sum_i^N \delta(\mathbf{x}-\mathbf{x}_i)$, $k$ is the modulus of the wavevector $\mathbf{k}$, and $\theta$ the angle of it in polar coordinates. The radially-averaged Fourier power will usually exhibit a peak at $k\neq0$, which we refer to as the characteristic wavenumber $k_c$ and relates to the characteristic spatial scale of the periodic pattern, $l_c$, by $l_c=2\pi / k_c$. 

\begin{figure}[ht!]
    \centering
    \includegraphics[width=0.75\columnwidth]{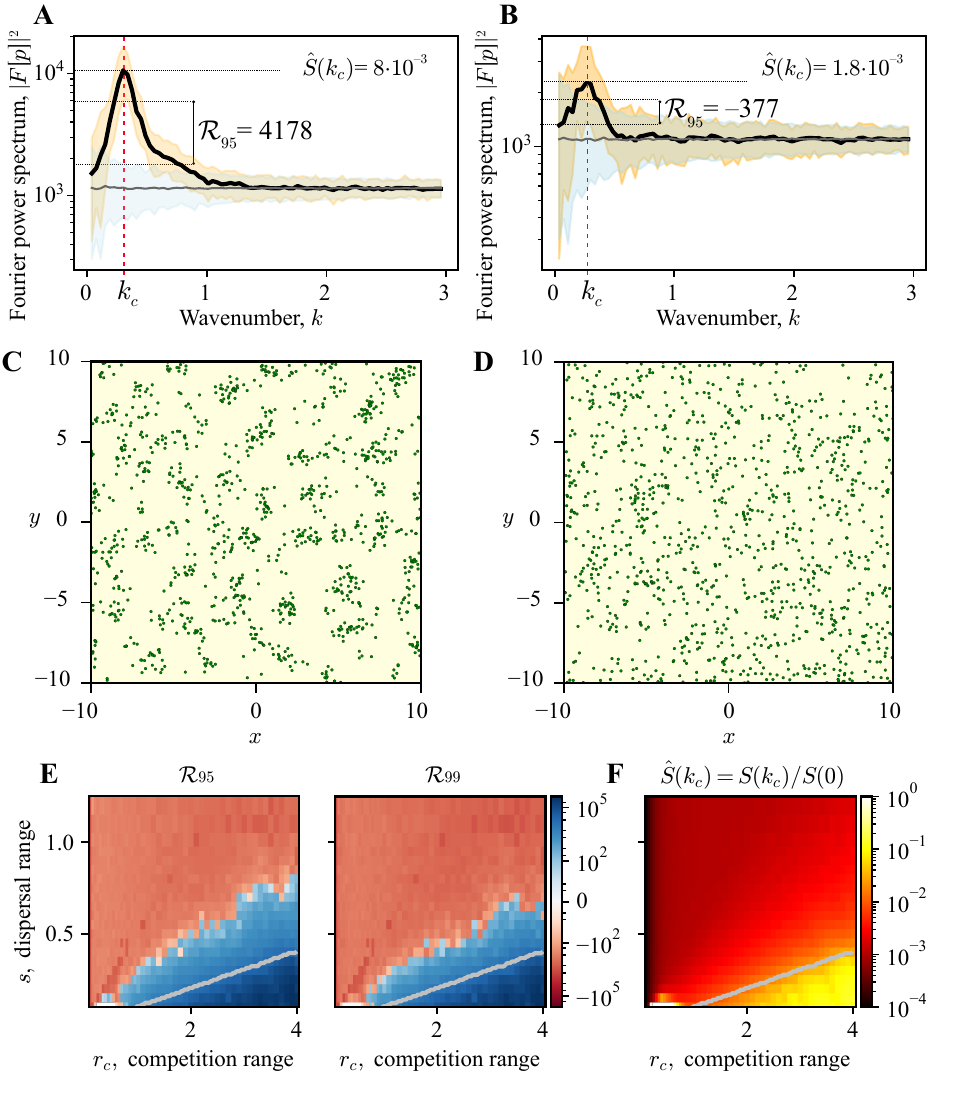}
    \caption{ Accuracy of the density field approximation predicting IBM-generated regular patterns. A,\,B) Radially-averaged Fourier power spectra distributions obtained from IBM simulations (black line) and spatially random point patterns (gray line). Envelopes to all curves represent the 2.5$\%$-97.5$\%$ interpercentile range. We computed power spectra of IBM-simulated patterns using 10 pattern snapshots for each of the 10 independent realizations we ran for each parameter combination (100 snapshots in total for each parameter set). For the randomized patterns, we randomized 100 times each of the 10 IBM realizations. Annotations in each panel indicate the value of the regularity metric, $\mathcal{R}_{95}>0$ for a regular pattern in A and $\mathcal{R}_{95}<0$ for non-regular pattern in B. C,\,D) Pattern snapshots used to compute the power spectra in A and B, respectively. Parameters: $s=0.3$ and $r_c=2.2$ (A, C); $s=0.65$ and $r_c=2.2$ (B, D). E) $\mathcal{R}_C$ metric for two different thresholds, $95\%$ and $99\%$ in the $r_c$, $s$ parameter space. F) Normalized power spectrum amplitude at the characteristic wavenumber $\hat{S}(k_c)=S(k_c)/S(0)$, as a function of the dispersal and competition spatial scales, $s$ and $r_c$. The gray line in E and F indicates the pattern formation onset as predicted by the density-based approximation.}
    \label{Fig:regularity metric}
\end{figure}

The first metric, regularity $\mathcal{R}_{C}$, compares the radially-averaged Fourier power spectra of IBM-simulated patterns with those of homogeneous distributions of individuals. To perform this comparison, we first generated an ensemble of spatial patterns for each parameter combination using IBM simulations and obtained their radially-averaged Fourier power spectrum. Then, we obtained the completely spatial random counterpart of each simulated pattern by randomizing the individual locations to be uniformly distributed within the simulation domain. We obtained the radially-averaged Fourier power spectra for these randomized patterns too. Lastly, we computed the mean and 2.5$\%$-97.5$\%$ interpercentile range of the power spectra distributions, both for simulated and randomized patterns. We define the regularity $\mathcal{R}_{95}$ as the signed distance between the 2.5$\%$ percentile of the IBM radially-averaged Fourier power spectrum distribution and the 97.5$\%$ percentile of its random counterpart, evaluated at the pattern characteristic wavenumber (i.e., the peak of the IBM radially-averaged Fourier power spectrum). The value of $\mathcal{R}_C$ for different values of $C$ is defined analogously. In Fig.\,\ref{Fig:regularity metric}, we show two examples in which this metric returns regular (panels A and C) and non-regular (panels B and D) patterns. We identify regular patterns as those associated with a positive regularity metric, $\mathcal{R}_C>0$, noting that the $C$\% interpercentile ranges do not overlap at the peak. Correspondingly, patterns with negative regularity metric ($\mathcal{R}_C<0$) are identified as non-regular by our metric, regardless of exhibiting a peak in the radially-averaged Fourier power spectrum. The reason is that, for $\mathcal{R}_C<0$, there is a non-negligible chance that a random pattern could exhibit the same peak in the Fourier power spectrum as a corresponding IBM simulation.

 The second metric, the relative power of the characteristic wavenumber $\hat{S}(k_c)$, quantifies the strength of periodicity in the IBM-simulated patterns by measuring the amplitude of the radially averaged Fourier power spectrum at the pattern wavelength relative to its value at $k=0$. We normalize this power by the power at $k=0$, which corresponds to $S(0)= N^2$, where $N$ is the total population size. Then, the second metric is defined as $\hat S(k_c)= S(k_c)/S(0)$. According to this metric, a perfectly linear periodic pattern (a sine function, for example) has all the power (for $k\neq0$) in the peak of the spectrum and would exhibit a value of $\hat{S}(k_c)=1$. On the other hand, a distribution with no periodicity shows a flat power spectrum (for $k\neq0$), and would exhibit a value $\hat{S}(k_c)\rightarrow 0$.

Next, we apply these two metrics to spatial patterns generated by IBM simulations and compare the results to the outcome of the linear stability analysis performed to the field equation in Eq.\,\eqref{lin-growth rate}. Both metrics identify periodic patterns in the IBM simulations when the dispersal range is short and the competition range is large (Fig.\,\ref{Fig:regularity metric}E,\,F). To compare these results with the predictions of density field equations, we evaluated the pattern formation onset by imposing the conditions for a Turing instability, $\lambda(\mathbf{k_c})=0$ and $\nabla \lambda (\mathbf{k_c}) =0$, in Eq.\,\eqref{lin-growth rate}. These equations define a curve in the parameter space for the threshold of a Turing instability (gray curves in Figs.\,\ref{Fig:regularity metric}E,\,F), which limits a region where the density equation predicts periodic patterns. This prediction identifies the region where IBM-generated patterns show the highest regularity and relative power in the characteristic wavenumber (bottom right corner in Figs.\,\ref{Fig:regularity metric}E,\,F). There are, however, parameter combinations for which the density field approximation does not capture the regular patterns returned by the IBM (blue regions on top of the gray line in Fig.\,\ref{Fig:regularity metric}E).
Incorporating demographic noise in the density equation could increase the agreement with IBM simulations, as noise can destabilize the homogeneous solution in parameter regimes where it is determinically stable \cite{Butler2009,Martinez-Garcia2013a}. Another possible strategy to refine the predictions of the density equation is to consider the dynamics of higher order correlations, similarly to the way they are treated in the SMD approximation, but explicitly in space.

\subsection{Comparison of the accuracy of both approximations.}
\begin{figure}[ht!]
	\centering
\includegraphics[width=\textwidth]{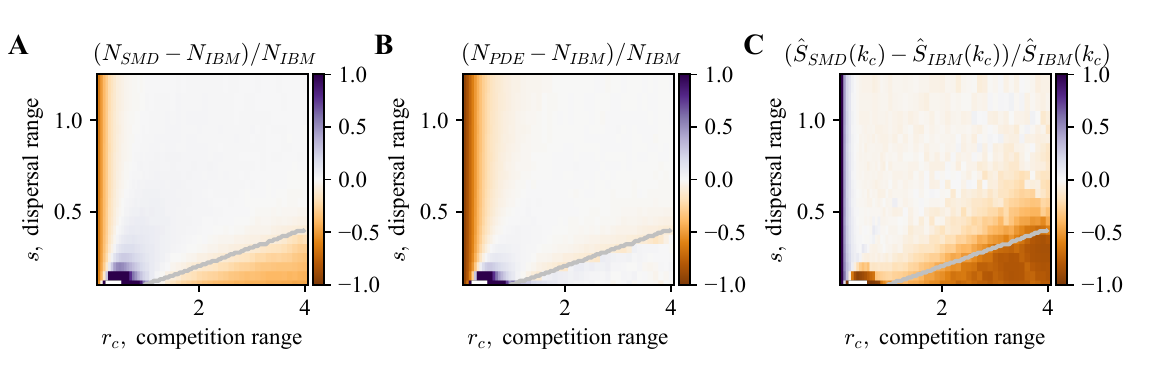}
	\caption{ Relative error of the PDE and SMD approximations compared to the IBM results for various competition and dispersal spatial scales, $r_c$ and $s$. A, B) Error in the total population size relative to IBM simulations using SMD (A) and density-based (B) approximations. C) Relative error in the relative amplitude of the characteristic wavenumber computed from SMD with the one obtained directly from IBM simulations. In all panels, the gray line indicates the onset of pattern formation as predicted by the density-based approximation.}
	\label{Fig:PDEcomparison}
\end{figure}

Our results so far show that spatial moment dynamics and density field PDEs provide complementary approximations to IBMs, with density field equations approximating IBMs more accurately than SMD in regions where highly regular periodic patterns form and SMD outcompeting density equations when patterns are not periodic but still not uniform. We quantified these different accuracies throughout the $(s,\,r_c)$ parameter space by computing the relative error of the density field and SMD equations at reproducing the total population size and spatial patterns from IBM simulations (Fig.\,\ref{Fig:PDEcomparison}). For population sizes, we defined the relative error in each approximation as
\begin{equation}
    \epsilon_\alpha = \frac{N_{\alpha}-N_\mathrm{IBM}}{N_\mathrm{IBM}},
\end{equation}
where $\alpha$ is a placeholder index for PDE (density equation) or SMD. Positive values of $\epsilon$ thus indicate that the deterministic approximation overestimates population size, while negative values indicate underestimation and $\epsilon=0$ a perfect agreement between the approximation and IBM simulations. To define a similar relative error in the characterization of the spatial patterns, we first need to write one of the Fourier-based regularity metrics in terms of quantities that can be measured from SMD analysis. We can write the Fourier power spectrum using the definition of the density field auto-correlation, $Z_2$
\begin{eqnarray}
    \lvert F[n]\rvert ^2 &=& \left| \int n(\mathbf{x}) e^{i\mathbf{k}\cdot\mathbf{x}} d\mathbf{x} \right| ^2, \nonumber \\
    &=& \int \int n(\mathbf{x})n(\mathbf{x'})e^{i\mathbf{k}\cdot(\mathbf{x-x')}}d\mathbf{x}d\mathbf{x'}, \nonumber \\
    &=& F\left[ \int n(\mathbf{x})n(\mathbf{x+x'})d\mathbf{x} \right], \nonumber \\
    &=& F [ A Z_2(\mathbf{x'}) + \int n(\mathbf{x})\delta(\mathbf{x'}) d\mathbf{x} ], \nonumber \\
    &=& A F[Z_2] + N,
\end{eqnarray}
where $A$ is the area and $N= AZ_1$ the population abundance. This derivation gives a relationship between $Z_2$, $Z_1$ and the Fourier power spectrum that we can evaluate using the solutions of the SMD equations to obtain the relative power $\hat{S}(k_c)$ and its relative error with respect to IBM-generated patterns.

We measured these relative errors in population size and spatial pattern structure and both for SMD and density-based approximations (Fig.\,\ref{Fig:PDEcomparison}). SMD underestimates the population size in the region where regular aggregated pattern formation occurs, while the density-based approximation is accurate in this regime (Figs.\,\ref{Fig:PDEcomparison}A\,,B). To explain this limitation of SMD, we computed the SMD relative error at reproducing the radially-averaged Fourer power spectrum of IBM-generated periodic patterns. Although SMD successfully captures short-distance correlations from IBM patterns (see measures of $C(0)$ in Fig.\,\ref{Fig:IBM-tophat}), it underestimates long-distance ones, which results in large relative errors in $\hat S(k_c)$ (Fig.\,\ref{Fig:PDEcomparison}C). This underestimation of long-range spatial correlations indicates that SMD predicts spatial patterns to be more uniform than those observed in IBM simulations, which ultimately makes SMD to also underestimate population size, as uniformly distributed populations have fewer individuals than those forming periodic patterns. Finally, both approximations underestimate population abundances---with SMD showing a smaller relative error---when short-range competition and long-range dispersal jointly lead to segregated spatial patterns.

\begin{figure}[ht!]
	\centering
\includegraphics[]{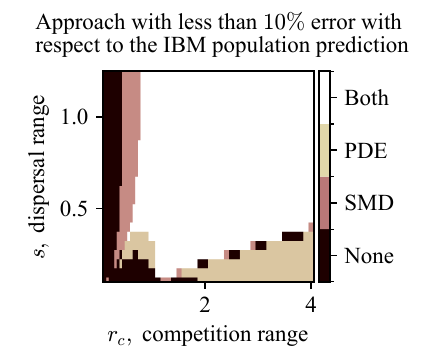}
	\caption{ Performance classification of SMD and density-based approximations when predicting population sizes from IBM simulations. Color code: white indicates that both the PDE and SMD approximations accurately match the IBM results; khaki denotes regions where only the PDE performs well; dark peach corresponds to regions where only the SMD is accurate; and black indicates parameter combinations where neither approximation is satisfactory. Performance is evaluated using the relative error in total population size, with a $10\%$ threshold.}
	\label{Fig:Summary}
\end{figure}

Finally, to summarize this systematic comparison between existing approximations to IBM spatial dynamics, we defined a scoring criterion based on how accurately they reproduce the population abundance of the IBM simulations. We set an error threshold of $10\%$ such that we considered SMD or density-based approximations to correctly reproduce IBM simulations if $\epsilon<0.1$. We computed $\epsilon$ for each combination of parameters $(r_c,\,s)$, identifying the regions of the parameter space where SMD and density equations outperform each other, both approximations are accurate or none of them provides a good description of the stochastic individual-based dynamics (Fig.\,\ref{Fig:Summary}). In general, the density-based approximation is accurate for high values of the competition range and low values of the dispersal range, while SMD provides better population size estimates for small competition ranges. However, when competition ranges become too small, SMD also starts to provide inaccurate predictions. Additionally, both methods perform poorly close to the origin, where populations are small and fluctuations are relevant. Finally, as expected, when competition and dispersal ranges are high, both approaches perform equally well. This occurs because the IBM simulations reach a \textit{well-mixed} regime, the population is randomly distributed, and spatial correlations are not relevant to set the population size.

\section{Conclusions}
We investigated the spatial patterns that form in a spatially explicit individual-based logistic model in which dispersal and intraspecific competition are long-range, using both a Gaussian and top-hat competition kernel. Combining numerical simulations of the discrete model, spatial moment dynamics and density field based approximations, we found that for top-hat competition kernels, the population might exhibit regular periodic patterns with aggregates of individuals forming a hexagonal lattice \cite{Lopez2004}. This type of pattern is not present when long-range competition is modeled using a Gaussian kernel \cite{law2003}, this fact is due to our IBM rules, and can be seen mathematically by inserting a Gaussian kernel in Eq.\,\eqref{lin-growth rate}; however, there are models for which a Gaussian competition kernel can produce patterns, highlighting that the issue of the general conditions for pattern formation is still debated \cite{MartinezGarcia2023}. Interestingly, the emergence of these regular patterns also affects the system productivity, allowing for larger population sizes even in conditions where local individual crowding is high \cite{Silvano2024}.

Next, we looked at how accurately a field equation for the biomass density, obtained by upscaling the individual-based dynamics, reproduced the existence of regular patterns. We found that the standard approximation in the coarse-graining procedure, which neglects spatial covariates in the density field, does not fully capture the region of the parameter space with regular patterns, being successful mainly in regions of the parameter space where patterns are more regular. Refining this approximation to account for demographic noise could significantly improve the agreement with IBMs. Noise modifies the power spectrum of the solutions of the density equation, enlarging the region in which the system displays a regular patterns \cite{Butler2009,Martinez-Garcia2013a}. On the other hand, the SMD equations poorly capture this region of the parameter space where periodic patterns form, underestimating the total population and the relative Fourier power spectrum amplitude at the characteristic wavenumber. Conversely, for most parameter combinations where the space-varying density approximation does not hold, SMD provides accurate predictions of the total population size and the spatial correlations. These results highlight the complementarity of these two approaches, each of them being the best approximation to IBM stochastic dynamics in particular regimes.

For ecological modeling applications in general and particularly in the case of vegetation patterns, individual-based models provide a robust and accurate framework to simulate individual-level processes and interactions, capturing dynamics down to the resolution of single plants. Consequently, these models are very effective for understanding the fine-scale mechanisms driving vegetation pattern formation. However, the computational overhead in simulating these IBMs becomes prohibitively expensive when the system size is large or when longer temporal scales are considered, limiting their practical applicability. Density field-based models, on the other hand, are computationally very efficient and often amenable to analytical exploration. However, these models assume that interactions are, to some extent, well-mixed, which leads to inaccuracies when trying to determine in which conditions spatial patterns form in vegetation systems. The SMD approximation does not resolve patterns explicitly and thus provides bad estimates of population sizes when highly regular and periodic patterns form. However, it retains key insights from IBMs regarding spatial patterns and spatially explicit interactions for parameter combinations where density equations provide inaccurate results. These results underscore that SMD and density-based approximations should be seen as complementary---rather than conflicting---approaches in the development of analytical approximations for spatially extended systems of interacting individuals.

\section*{Acknowledgments}
This work was partially funded by the Center of Advanced Systems Understanding (CASUS), which is financed by Germany’s Federal Ministry of Education and Research (BMBF) and by the Saxon Ministry for Science, Culture and Tourism (SMWK) with tax funds on the basis of the budget approved by the Saxon State Parliament. RMG was supported by FAPESP through grant ICTP-SAIFR 2021/14335-0. RM was supported by CNPq through grant 140096/2021-3 and CAPES - Finance Code 001.

\bibliography{references}

\begin{thebibliography}{10}

\bibitem{Pringle2017}
Pringle RM, Tarnita CE.
\newblock Spatial {Self}-{Organization} of {Ecosystems}: {Integrating}
  {Multiple} {Mechanisms} of {Regular}-{Pattern} {Formation}.
\newblock Annual Review of Entomology. 2017;62(1):359-77.

\bibitem{Lee2021}
Lee ED, Kempes CP, West GB.
\newblock Growth, death, and resource competition in sessile organisms.
\newblock Proceedings of the National Academy of Sciences of the United States
  of America. 2021;118(15).
\newblock ISBN: 2020424118.

\bibitem{Rietkerk2008}
Rietkerk M, van~de Koppel J.
\newblock Regular pattern formation in real ecosystems.
\newblock Trends in ecology \& evolution. 2008 Mar;23(3):169-75.
\newblock Publisher: Elsevier Science Publishers.
\newblock Available from:
  \url{http://linkinghub.elsevier.com/retrieve/pii/S0169534708000281}.

\bibitem{Martinez-Garcia2022}
Martinez-Garcia R, Tarnita CE, Bonachela JA.
\newblock Self-organized patterns in ecological systems: from microbial
  colonies to landscapes.
\newblock Emerging Topics in Life Sciences. 2022;6(3):245-58.
\newblock Available from:
  \url{https://portlandpress.com/emergtoplifesci/article-abstract/6/3/245/231411/Spatial-patterns-in-ecological-systems-from?redirectedFrom=fulltext}.

\bibitem{Tarnita2024}
Tarnita CE.
\newblock Self-organization in spatial ecology.
\newblock Current Biology. 2024 Oct;34(20):R965-70.
\newblock Available from:
  \url{https://linkinghub.elsevier.com/retrieve/pii/S0960982224012430}.

\bibitem{Rohani1997}
Rohani P, Lewis TJ, Grünbaum D, Ruxton GD.
\newblock Spatial self-organization in ecology: {Pretty} patterns or robust
  reality?
\newblock Trends in Ecology and Evolution. 1997;12(2):70-4.

\bibitem{Dong2019}
Dong X, Fisher SG.
\newblock Ecosystem spatial self-organization: {Free} order for nothing?
\newblock Ecological Complexity. 2019 Apr;38:24-30.
\newblock Available from:
  \url{https://linkinghub.elsevier.com/retrieve/pii/S1476945X18301508}.

\bibitem{rietkerk2021evasion}
Rietkerk M, Bastiaansen R, Banerjee S, van~de Koppel J, Baudena M, Doelman A.
\newblock Evasion of tipping in complex systems through spatial pattern
  formation.
\newblock Science. 2021;374(6564):eabj0359.

\bibitem{Kefi2024}
K{\'e}fi S, G{\'e}nin A, Garcia-Mayor A, Guirado E, Cabral JS, Berdugo M,
  et~al.
\newblock Self-organization as a mechanism of resilience in dryland ecosystems.
\newblock Proceedings of the National Academy of Sciences.
  2024;121(6):e2305153121.

\bibitem{Rietkerk2002}
Rietkerk M, Boerlijst MC, van Langevelde F, HilleRisLambers R, de~Koppel Jv,
  Kumar L, et~al.
\newblock Self‐{Organization} of {Vegetation} in {Arid} {Ecosystems}.
\newblock The American Naturalist. 2002 Oct;160(4):524-30.
\newblock ISBN: 0003-0147.

\bibitem{Lefever1997}
Lefever R, Lejeune O.
\newblock On the origin of tiger bush.
\newblock Bulletin of Mathematical biology. 1997;59:263-94.

\bibitem{Deblauwe2011}
Deblauwe V, Couteron P, Lejeune O, Bogaert J, Barbier N.
\newblock Environmental modulation of self-organized periodic vegetation
  patterns in {Sudan}.
\newblock Ecography. 2011;34(6):990-1001.

\bibitem{Clerc2021}
Clerc MG, Echeverr{\'\i}a-Alar S, Tlidi M.
\newblock Localised labyrinthine patterns in ecosystems.
\newblock Scientific reports. 2021;11(1):18331.

\bibitem{Guirado2023}
Guirado E, Delgado-Baquerizo M, Benito BM, Molina-Pardo JL, Berdugo M,
  Mart{\'\i}nez-Valderrama J, et~al.
\newblock The global biogeography and environmental drivers of fairy circles.
\newblock Proceedings of the National Academy of Sciences.
  2023;120(40):e2304032120.

\bibitem{Gandhi2018}
Gandhi P, Werner L, Iams S, Gowda K, Silber M.
\newblock A topographic mechanism for arcing of dryland vegetation bands.
\newblock Journal of The Royal Society Interface. 2018;15(147):20180508.
\newblock Available from:
  \url{http://dx.doi.org/10.1098/rsif.2018.0508Electronicsupplementarymaterialisavailableonlineathttps://dx.doi.org/10.6084/m9.figshare.c.4238546.}

\bibitem{Gowda2018}
Gowda K, Iams S, Silber M.
\newblock Signatures of human impact on self-organized vegetation in the {Horn}
  of {Africa}.
\newblock Scientific Reports. 2018;8(1):1-8.
\newblock ArXiv: 1705.05308 Publisher: Springer US.
\newblock Available from: \url{http://dx.doi.org/10.1038/s41598-018-22075-5}.

\bibitem{Siteur2023}
Siteur K, Liu QX, Rottschäfer V, Heide Tvd, Rietkerk M, Doelman A, et~al.
\newblock Phase-separation physics underlies a new theory for the resilience of
  patchy ecosystems.
\newblock Proceedings of the National Academy of Sciences.
  2023;120(2):e2202683120.
\newblock ISBN: 1211925110.
\newblock Available from:
  \url{http://www.pnas.org/lookup/suppl/doi:10.1073/pnas.2216830120/-/DCSupplemental.https://doi.org/10.1073/pnas.2216830120}.

\bibitem{Pinto-Ramos2023}
Pinto-Ramos D, Clerc MG, Tlidi M.
\newblock Topological defects law for migrating banded vegetation patterns in
  arid climates.
\newblock Science Advances. 2023;9(31):1-12.

\bibitem{Hidalgo2024}
Hidalgo-Ogalde B, Pinto-Ramos D, Clerc MG, Tlidi M.
\newblock Nonreciprocal feedback induces migrating oblique and horizontal
  banded vegetation patterns in hyperarid landscapes.
\newblock Scientific reports. 2024;14(1):14635.

\bibitem{Bastiaansen2018}
Bastiaansen R, Jaïbi O, Deblauwe V, Eppinga MB, Siteur K, Siero E, et~al.
\newblock Multistability of model and real dryland ecosystems through spatial
  self-organization.
\newblock Proceedings of the National Academy of Sciences. 2018
  Oct;115(44):11256-61.
\newblock Available from:
  \url{http://www.pnas.org/lookup/doi/10.1073/pnas.1804771115}.

\bibitem{Veldhuis2022}
Veldhuis MP, Martinez-Garcia R, Deblauwe V, Dakos V.
\newblock Remotely-sensed slowing down in spatially patterned dryland
  ecosystems.
\newblock Ecography. 2022;10:e06139.

\bibitem{Lejeune1999}
Lejeune O, Tlidi M.
\newblock A model for the explanation of vegetation stripes (tiger bush).
\newblock Journal of Vegetation Science. 1999 Apr;10(2):201-8.
\newblock Publisher: Blackwell Publishing Ltd.
\newblock Available from: \url{http://dx.doi.org/10.2307/3237141}.

\bibitem{Meron2004}
Meron E, Gilad E, Von~Hardenberg J, Shachak M, Zarmi Y.
\newblock Vegetation patterns along a rainfall gradient.
\newblock Chaos, Solitons and Fractals. 2004;19(2):367-76.
\newblock ISBN: 0960-0779.

\bibitem{Kealy2012}
Kealy BJ, Wollkind DJ.
\newblock A {Nonlinear} {Stability} {Analysis} of {Vegetative} {Turing}
  {Pattern} {Formation} for an {Interaction}–{Diffusion} {Plant}-{Surface}
  {Water} {Model} {System} in an {Arid} {Flat} {Environment}.
\newblock Bulletin of Mathematical Biology. 2012 Apr;74(4):803-33.

\bibitem{MartinezGarcia2013}
Martinez-Garcia R, Calabrese JM, Hernández-García E, López C.
\newblock Vegetation pattern formation in semiarid systems without facilitative
  mechanisms.
\newblock Geophysical Research Letters. 2013;40:6143-7.

\bibitem{PintoRamos2024}
Pinto-Ramos D, Clerc MG, Makhoute A, Tlidi M.
\newblock Vegetation clustering and self-organization in inhomogeneous
  environments.
\newblock arXiv. 2024 Jun.
\newblock 2406.12581.
\newblock Available from: \url{http://arxiv.org/abs/2406.12581}.

\bibitem{Durrett1994}
Durrett R, Levin SA.
\newblock The importance of being discrete (and spatial).
\newblock Theoretical Population Biology. 1994;46:363-94.
\newblock ArXiv: 1305.0446v1.
\newblock Available from: \url{http://arxiv.org/abs/1305.0446}.

\bibitem{Bolker1997}
Bolker BM, Pacala SW.
\newblock Using moment equations to understand stochastically driven spatial
  pattern formation in ecological systems.
\newblock Theoretical Population Biology. 1997;52(3):179-97.

\bibitem{Wiegand2021}
Wiegand T, Wang X, Anderson-Teixeira KJ, Bourg NA, Cao M, Ci X, et~al.
\newblock Consequences of spatial patterns for coexistence in species-rich
  plant communities.
\newblock Nature Ecology \& Evolution. 2021;5(7):965-73.

\bibitem{Bolker1999}
Bolker BM, Pacala SW.
\newblock Spatial moment equations for plant competition: understanding spatial
  strategies and the advantages of short dispersal.
\newblock The American Naturalist. 1999;153(6):575-602.

\bibitem{MartinezGarcia2023}
Martinez-Garcia R, Cabal C, Calabrese JM, Hern{\'a}ndez-Garc{\'\i}a E, Tarnita
  CE, L{\'o}pez C, et~al.
\newblock Integrating theory and experiments to link local mechanisms and
  ecosystem-level consequences of vegetation patterns in drylands.
\newblock Chaos, Solitons \& Fractals. 2023;166:112881.

\bibitem{Borgogno2009}
Borgogno F, D'Odorico P, Laio F, Ridolfi L.
\newblock Mathematical models of vegetation pattern formation in ecohydrology.
\newblock Reviews of Geophysics. 2009;47(1).
\newblock Available from: \url{http://dx.doi.org/10.1029/2007RG000256}.

\bibitem{Zelnik2017}
Zelnik YR, Uecker H, Feudel U, Meron E.
\newblock Desertification by front propagation?
\newblock Journal of Theoretical Biology. 2017;418:27-35.

\bibitem{law2003}
Law R, Murrell DJ, Dieckmann U.
\newblock Population {Growth} in {Space} and {Time} : {Spatial} {Logistic}
  {Equations}.
\newblock Ecology. 2003;84(1):252-62.

\bibitem{sheffer2012}
Sheffer E, Yizhaq H, Meron E.
\newblock Emerged or imposed: a theory on the role of physical templates and
  self‐organisation for vegetation patchiness.
\newblock Ecology Letters. 2012.

\bibitem{yizhaq2014}
Yizhaq H, Sela S, Svoray T, Assouline S, Bel G.
\newblock Effects of heterogeneous soil‐water diffusivity on vegetation
  pattern formation.
\newblock Water Resources Research. 2014.

\bibitem{echeverria2023effect}
Echeverr{\'\i}a-Alar S, Pinto-Ramos D, Tlidi M, Clerc M.
\newblock Effect of heterogeneous environmental conditions on labyrinthine
  vegetation patterns.
\newblock Physical Review E. 2023;107(5):054219.

\bibitem{pinto2022vegetation}
Pinto-Ramos D, Echeverr{\'\i}a-Alar S, Clerc MG, Tlidi M.
\newblock Vegetation covers phase separation in inhomogeneous environments.
\newblock Chaos, Solitons \& Fractals. 2022;163:112518.

\bibitem{Butler2009}
Butler T, Goldenfeld N.
\newblock Robust ecological pattern formation induced by demographic noise.
\newblock Physical Review E. 2009 Sep;80(3):030902.
\newblock Publisher: American Physical Society.
\newblock Available from:
  \url{http://link.aps.org/doi/10.1103/PhysRevE.80.030902}.

\bibitem{Martinez-Garcia2013a}
Martinez-Garcia R, Calabrese JM, López C.
\newblock Spatial patterns in mesic savannas: {The} local facilitation limit
  and the role of demographic stochasticity.
\newblock Journal of Theoretical Biology. 2013 Jun;333(0):156-65.
\newblock ArXiv: 1209.5178v4 Publisher: Elsevier.
\newblock Available from: \url{http://dx.doi.org/10.1016/j.jtbi.2013.05.024}.

\bibitem{Dasilva2014}
Da~Silva LA, Colombo EH, Anteneodo C.
\newblock Effect of environment fluctuations on pattern formation of single
  species.
\newblock Physical Review E - Statistical, Nonlinear, and Soft Matter Physics.
  2014;90(1):1-8.

\bibitem{Karig2018}
Karig D, Martini KM, Lu T, DeLateur NA, Goldenfeld N, Weiss R.
\newblock Stochastic {Turing} patterns in a synthetic bacterial population.
\newblock Proceedings of the National Academy of Sciences. 2018:201720770.
\newblock ArXiv: 1408.1149 ISBN: 1720770115.
\newblock Available from:
  \url{http://www.pnas.org/lookup/doi/10.1073/pnas.1720770115}.

\bibitem{Surendran2020}
Surendran A, Plank MJ, Simpson MJ.
\newblock Population dynamics with spatial structure and an {Allee} effect.
\newblock Proceedings of the Royal Society A: Mathematical, Physical and
  Engineering Sciences. 2020;476:1-19.

\bibitem{Rogers2019}
Rogers HS, Beckman NG, Hartig F, Johnson JS, Pufal G, Shea K, et~al.
\newblock The Total Dispersal Kernel: A Review and Future Directions.
\newblock AoB PLANTS. 2019 Sep;11(5):plz042.

\bibitem{Bullock2017}
Bullock JM, Mallada~Gonz{\'a}lez L, Tamme R, G{\"o}tzenberger L, White SM,
  P{\"a}rtel M, et~al.
\newblock A Synthesis of Empirical Plant Dispersal Kernels.
\newblock Journal of Ecology. 2017;105(1):6-19.

\bibitem{A_Surendran_2020}
Surendran A, Plank MJ, Simpson MJ.
\newblock Small-scale spatial structure affects predator-prey dynamics and
  coexistence.
\newblock Theoretical Ecology. 2020;13:537-50.

\bibitem{Lopez2004}
López C, Hernández-García E.
\newblock Fluctuations impact on a pattern-forming model of population dynamics
  with non-local interactions.
\newblock Physica D: Nonlinear Phenomena. 2004.

\bibitem{Silvano2024}
Silvano NO, Valeriano J, Hernández-García E, López C, Martinez-Garcia R.
\newblock Shear and transport in a flow environment determine spatial patterns
  and population dynamics in a model of nonlocal ecological competition.
\newblock arXiv. 2024 Sep.
\newblock 2409.04268.
\newblock Available from: \url{http://arxiv.org/abs/2409.04268}.

\bibitem{Hirsch_2012}
Hirsch BT, Visser MD, Kays R, Jansen PA.
\newblock Quantifying seed dispersal kernels from truncated seed-tracking data.
\newblock Methods in Ecology and Evolution. 2012;3(3):595-602.
\newblock Available from:
  \url{https://besjournals.onlinelibrary.wiley.com/doi/abs/10.1111/j.2041-210X.2011.00183.x}.

\bibitem{Cabal_2020}
Cabal C, Martínez-García R, Aguilar A, Valladares F, Pacala SW.
\newblock The exploitative segregation of plant roots.
\newblock Science. 2020;370(6521):1197-9.
\newblock Available from:
  \url{https://www.science.org/doi/abs/10.1126/science.aba9877}.

\bibitem{Plank2015}
Plank MJ, Law R.
\newblock Spatial {Point} {Processes} and {Moment} {Dynamics} in the {Life}
  {Sciences}: {A} {Parsimonious} {Derivation} and {Some} {Extensions}.
\newblock Bulletin of Mathematical Biology. 2015 Apr;77(4):586-613.
\newblock Available from:
  \url{http://link.springer.com/10.1007/s11538-014-0018-8}.

\bibitem{Surendran2021}
Surendran A.
\newblock Stochastic and continuum descriptions of population dynamics with
  spatial structure [PhD thesis].
\newblock Queensland University of Technology; 2021.
\newblock Available from: \url{https://eprints.qut.edu.au/207574/}.

\bibitem{Surendran2018}
Surendran A, Plank MJ, Simpson MJ.
\newblock Spatial moment description of birth–death–movement processes
  incorporating the effects of crowding and obstacles.
\newblock Bulletin of Mathematical Biology. 2018;80:2828-55.

\bibitem{Surendran2019}
Surendran A, Plank MJ, Simpson MJ.
\newblock Spatial structure arising from chase-escape interactions with
  crowding.
\newblock Scientific Reports. 2019;9:14988.

\bibitem{Murrell_et_al_2004}
Murrell DJ, Dieckmann D, Law R.
\newblock On moment closures for population dynamics in continuous space.
\newblock Journal of theoretical biology. 2004;229:421-32.

\bibitem{Raghib_et_al_2011}
Raghib M, Hill NA, Dieckmann U.
\newblock A multiscale maximum entropy moment closure for locally regulated
  space–time point process models of population dynamics.
\newblock Journal of Mathematical Biology. 2011;62:605-53.

\bibitem{Binny2016}
Binny RN, James A, Plank MJ.
\newblock Collective cell behaviour with neighbour-dependent proliferation,
  death and directional bias.
\newblock Bulletin of Mathematical Biology. 2016;78:2277-301.

\bibitem{Law_and_Dieckmann_2000}
Law R, Dieckmann U.
\newblock A dynamical system for neighborhoods in plant communities.
\newblock Ecology. 2000;81:2137-48.

\bibitem{van1992stochastic}
Van~Kampen NG.
\newblock Stochastic processes in physics and chemistry. vol.~1.
\newblock Elsevier; 1992.

\bibitem{Toral2014}
Toral R, Colet P.
\newblock Stochastic numerical methods: an introduction for students and
  scientists.
\newblock John Wiley \& Sons; 2014.

\bibitem{Murray2007ch11}
Murray JD.
\newblock Reaction Diffusion, Chemotaxis, and Nonlocal Mechanisms.
\newblock In: Mathematical biology: I. An introduction. vol.~17. Springer
  Science \& Business Media; 2007. p. 395-416.

\bibitem{Jorge2024}
Jorge DCP, Martinez-Garcia R.
\newblock Demographic effects of aggregation in the presence of a component
  {Allee} effect.
\newblock Journal Royal Society Interface. 2024 Jun;21:20240042.
\newblock Available from:
  \url{https://royalsocietypublishing.org/doi/10.1098/rsif.2024.0042}.

\bibitem{meron2015nonlinear}
Meron E.
\newblock Nonlinear physics of ecosystems.
\newblock CRC Press, Taylor \& Francis Group Boca Raton, FL, USA; 2015.

\end{thebibliography}

\end{document}